\documentclass[11pt,twoside]{book}
\usepackage{newpasp}

\begin{document}

\setcounter{page}{659}
\markboth{Hibbard et al.}{An \HI\ Rogues Gallery}

\title{An \HI\ Rogues Gallery}
\author{J.E.\ Hibbard}
\affil{NRAO}
\author{J.H.\ van Gorkom}
\affil{Columbia University}
\author{Michael P.\ Rupen}
\affil{NRAO}
\author{David Schiminovich}
\affil{Caltech}

\begin{abstract}
We have begun a compilation of \HI\ maps of peculiar galaxies kindly
contributed by individual investigators, many as part of the ``Gas \&
Galaxy Evolution'' conference. In this gallery we present images of
the first $\sim$180 objects, which includes over 400 individually 
cataloged galaxies. The images consist of a greyscale
representation of the optical morphology and an accompanying optical
image with \HI\ contours superimposed. A web-based ``living Gallery''
is being maintained on the NRAO homepage (presently at
{\it http://www.nrao.edu/astrores/HIrogues/}).
\end{abstract}

\section{Introduction}

Arp's ``Atlas of Peculiar Galaxies'' (Arp 1966) appeared forty years
after the discovery that galaxies are independent stellar systems.  It
was motivated by some of the questions which form central themes of
this conference: How are elliptical galaxies related to spirals?  How
are galaxies formed, and how do they evolve? These questions are just
as compelling today as they were in 1966. It is certainly true that
optical compilations of the bizarre objects collected by Arp and
others (notably, Vorontsov-Vel'yaminov 1959, 1977; Arp \& Madore 1987;
Malin \& Carter 1983) and the detailed studies that they motivated
have brought us an appreciation that peculiar galaxies have an
importance far in excess of their statistical occurrence in galaxy
catalogs, where they comprise anywhere from 3--9\% of all objects (Arp
\& Madore 1975; Struck 1999).  Still, thirty-five years later
these questions remain largely unanswered.

By the time atlases of peculiar galaxies appeared, there were several
catalogs devoted to demonstrating the various forms of galaxy
morphology. Most prominent among these is the Hubble Atlas. These
catalogs are arranged in sequences based on the symmetry of the
optical morphologies, and focus on single galaxies in isolation, with
asymmetric forms relegated to catch-all categories and illustrated
with but a few token examples. The compendiums of peculiar galaxies,
on the other hand, highlight those forms which deviate from symmetry,
with the thinking that such objects are transitional forms
illustrating the most spectacular phases of galactic evolution, and
perhaps formation.

Optical morphology alone provides an interesting yet incomplete view
of galaxies. This is particularly true when considering peculiar
systems. The starlight in galaxies rarely extends beyond a few optical
scale-lengths, well within the dark matter halos that are widely
believed to surround galaxies. Any tracer that extends to larger radii
will probe the influence of a larger fraction of the mass of galaxies
and, because the dynamical time scales with radius, will sample
galactic influences further in the past. The 21\,cm line of neutral
hydrogen forms just such a tracer.  Since the early days of radio
astronomy, it has been known that the neutral gas often extends much
further than the optical light (Roberts \& Rots 1973). Additionally,
hydrogen is the raw material from which stars and therefore galaxies
are ultimately formed, so if large reservoirs of it exist at the
present time, we might hope to find signatures of on-going formation
processes in the neutral gas.  No picture of galaxy formation and
evolution is complete without knowledge of how the gas is distributed
at large radii.

It is somewhat surprising, then, that thirty-five years after the
optical compilation of Arp, no such compilation of \HI\ maps exist for
peculiar galaxies. This is not for lack of such observations. \HI\
mapping requires an extensive investment in telescope time and
post-observation reduction, limiting most studies to a few to a dozen
objects in any give observational program. Nonetheless, an impressive
amount of data on \HI\ in weird galaxies and weird \HI\ in otherwise
normal galaxies has been amassed over the years. By our count, more
than 400 peculiar systems have been mapped in \HI. This may be far
below the number of objects typically assembled in optical catalogs,
but it is enough to warrant an attempt to assemble and arrange them in
some organized form.

This Gallery is intended to do just that, and we have taken the
occasion of this meeting, celebrating the twentieth anniversary of the
VLA, to make a first attempt at its compilation. We concentrate on
\HI\ maps of peculiar galaxies or peculiar \HI\ in otherwise normal
galaxies, rather than attempting to include all Hubble Types. We have
further limited ourselves to very disturbed systems, by and large
including only a few (or no) examples of systems with less extreme
peculiarities. There are already relatively large compilations of the
\HI\ morphology of ``normal'' Hubble types, many of which exhibit both
minor and major peculiarities, in (motsly Dutch) theses or journal
supplements (e.g. Wevers, van der Kruit \& Allen 1986; Bosma 1978;
Warmels 1988; Cayatte et al.\ 1990; van Driel \& van Woerden 1991;
Oosterloo 1988; Broeils 1992; Swaters 1999; Verheijen \& Sancisi
2001), and the growing database of the Westerbork \HI\ Survey of
Irregular and Spiral Galaxies (WHISP) will eventually contain more
than 400 late-type galaxies (van der Hulst, van Albada \& Sancisi,
these proceedings, p.~451).  On the other hand, no
large compilation of \HI\ in peculiars exits anywhere. This ``Rogues
Gallery'' is our attempt to remedy this situation.

This is a very subjective compilation, assembled because one of us
noticed the \HI\ map of a given object somewhere and considered it
interesting, or had marked the galaxy as an optically interesting
object and subsequently searched the literature or telescope archives
for existing \HI\ observations. This assemblage is definitely not
complete, and may not even be representative of all the forms of
peculiar \HI\ morphologies --- for that we would need an optically
blind \HI\ survey.  But one has to start somewhere, and this is where
we have decided to start.

Following the practice of Hubble, Sandage, Arp, and Madore, we have
divided the Gallery into several morphological classes. These are
described in detail in \S3. Within each class we have attempted to
arrange objects in a suggestive morphological sequence, where the
principles behind those sequences are described in \S3.1. The general
theme is the connection of similar forms, related from a dyamical
(as with the class of interacting galaxies) or a morphological viewpoint (as
with the class of galaxies with warps and/or asymmetries). Sometimes
there is no clear transition between objects, and the category is just
a compilation of systems that fall within the broadly defined class
(as with the class of galaxies with extended \HI\ envelopes).

The Gallery is laid out as follows. We describe the
subjective gathering of the images and the proper way to reference
them in \S2. In \S3 we define the different classes used and explain
how objects are ordered in the Gallery. The layout of the images is
described in \S4, while \S5 describes the tables and their entries.
The acknowledgements appear in \S6, followed by the tables, and finally the
Gallery itself. At the end of the Gallery we
have collected abstracts for authors who have allowed us to reproduce
images prior to their publication in the literature.

\section{Sources of Images and References}\label{sec:sources}

A list of candidate objects for the Gallery has been continuously
assembled from the literature and personal communications by
JEH. Prior to this meeting, we also searched the VLA archive and the
compilation of Martin (1998) for \HI\ mapping observations of galaxies
noted by us to be optically peculiar in some manner. As such, this is
an entirely subjective assemblage and can in no way be considered
complete. We do hope that it is representative (although it may well
not be).

Using this list as a guide, we tried to contact the observers who made
the observations in the early part of 2000 to see if they would
consider donating FITS images or postscript files of the resulting
\HI\ maps for inclusion in the Gallery. We made another attempt to
increase the number of objects in the early part of 2001. The present
collection includes the results (please see \S6 for acknowledgments). 
It amounts to 181 systems (including over 400 individually cataloged
galaxies) out of our original list of $\sim400$; there is thus much
room for improvement. Additionally, there are likely a large number of
objects, either published or unpublished, that we have missed
entirely. We plan on actively maintaining this Gallery on the NRAO
webpage (currently stored at {\it http://www.nrao.edu/astrores/HIrogues/}), 
and encourage observers who notice any omissions (or wish themselves to 
contribute an image) to contact us.

The \HI\ maps in this compilation are mainly from two major centimeter
wavelength interferometers, the Very Large Array (VLA) and the
Westerbork Synthesis Radio Telescope (WSRT).  A few images were taken
from observations with the Australia Telescope Compact Array (ATCA),
but there are undounbtedly dozens of systems observed with that
telescope which are not included here. Our hope is that we can remedy
this in the near future, or perhaps that there will be a Northern and
Southern Gallery maintained at NRAO and ATNF (a list of all galaxies
observed with the ATCA is presently maintained at {\it
http://wwwatnf.atnf.csiro.au/people/bkoribal/atca\_obs.html}).
Similarly, although no mapping observations from the Giant Metrewave
Radio Telescope (GMRT) are included so far, that instrument is likely
to become another major source of images in the future. It is notable
that we have included only four mapping observations made with single
dish telescopes, primarily because such maps are often unavailable in
machine readable form. We can expect very interesting results on
the large scale gaseous distribution of local galaxies from single
dish maps from the HIPASS survey with Parkes telescope (see Koribalski, 
these proceedings, p.~439) and the Green Bank Telescope 
in the near future.

Many of the images presented in this Gallery have appeared in the
literature. However, we are particularly and deeply grateful for those
observers who have made their data available prior to publication. To
ensure that these observers receive proper credit for this work, we
have collected short abstracts under their names at the end of the
Gallery, which include brief descriptions of these observations.
If readers wish to reference these images, they should cite the
abstracts, which are also referenced in the figure captions.

\section{Rogues Gallery Classifications}\label{sec:classes}

In this first attempt at constructing an atlas of this sort, we have
arranged the systems following the lead of Arp (1966). In particular,
we define several broad morphological classes, but make no
clear breaks between the classes, and the objects are often arranged
so that the last objects of one class are similar to the first
objects of the following one. Overall, the major motivation for
placing an object on a specific page is to have it appear between
systems that it most resembles.

These classes are defined morphologically, and within each category,
objects are presented in a suggestive sequence. The order within
a class is sometimes motivated by dynamical considerations, e.g., the
arrangement of interacting galaxies in a suggested evolutionary sequence.
These motivations and guiding principles are explained in the following
sub-section, which also gives detailed motivations for placing some specific
objects in one sequence rather than another.

Table~1 shows the arrangement of the Gallery, along with the criteria
for ordering of objects within classes, and Table~2 lists the
individual galaxies in the order in which they appear. As mentioned
above the last object of one class is often similar to the first
object of the next one.  The classification of galaxies is a very
subjective task, and many systems could easily fit into several
categories. This is a work in progress and we welcome your comments.

\subsection{Detailed Notes on Classifications and Ordering}
\label{sec:subclasses}


The first class of objects in the Gallery is {\bf Galaxies with
Extended \HI\ Envelopes}, where the \HI\ extent is compared to the
underlying optical light rather than to a fixed physical scale.
The rest of the Gallery is dominated by interacting systems, many of which also
have very extended \HI\ distributions. In this first class, we include
objects which do not {\it apparently} owe their extended \HI\ to
an interaction\footnote{This is a somewhat circular argument,
since the prime evidence for an interaction is often the extent of the gas
distribution, and there are papers in the literature for nearly all of
the objects in this class claiming an interacting or merging origin
based on this fact.}. By this we mean that there is not an obvious
partner with which the system is interacting, nor are there tell-tale
signs of a prior minor or major merger event (see below), such as
stellar shells and loops or multiple nuclei.

This classification includes many Blue Compact Dwarfs (BCDs), although
BCDs also appear elsewhere in the Gallery.  For example,
UGC\,00521 appears with warps, and there are
several BCDs filed under {\it Miscellaneous}, which may not have been mapped
with enough sensitivity to determine whether or not they have extended gas
envelopes. BCDs are some of the most unevolved systems in the
nearby universe, in terms of their stellar content and metallicity,
and as such this {\it Extended \HI}\ class may include examples of galaxies
actually
forming out of the gaseous reservoirs in which they are embedded. For
this reason, we have arranged galaxies within this class based on
the regularity of their \HI\ morphology: galaxies with very irregular
outer \HI\ are at the beginning, and galaxies with disk-like \HI\
distributions are at the end.

The next class is {\bf Galaxies with \HI\ Extensions}. The first subclass
is {\bf Galaxies with Two-Sided Warps}. This is the perhaps the most
under-represented class included in the Gallery, as warps are rather common
(Bosma 1991). Here we have simply collected some representative examples.
This includes the very peculiar system NGC\,3718
(Fig.~13).  We have placed it here because, like the
two-sided warps, it has a 180$^\circ$ sense of symmetry, and we could
not think of a more suitable place for it.

The two-sided warps are followed by {\bf Galaxies with One-Sided
\HI\ Extensions}. Within this subclass are galaxies whose \HI\ morphology
is believed to be due to Ram Pressure Stripping
(Figs.~21 --- 26). The next system,
the BCD galaxy NGC\,4861 (Fig.~27), has both an
extended \HI\ envelope and an \HI\ cloud with no optical counterpart,
making it the perfect connector to the next subclass, {\bf Galaxies
with Detached \HI\ Clouds}\footnote{We note that many of the galaxies
in other parts of the Gallery have \HI\ clouds with no identified
optical counterparts, but are classified according to their most
dominant \HI\ characteristics.}.  This is another group of galaxies
which is poorly represented in the Gallery, and many more examples
appear in the literature. The qualifier ``no optical counterpart''
obviously depends on the quality of the existing optical data, and to
hedge our bets, we have placed this classification between ``\HI\
extensions'' and the next class of systems, the {\it Minor
Mergers}. This way, if an optical counterpart to the \HI\ cloud is
subsequently identified, the Gallery needs not be re-ordered.
Similarly, the first member of the Minor Mergers,
M\,108 (Fig.~30), has an \HI\
extension toward an optical companion, but the optical companion does
not have a known redshift. If it is subsequently shown to be a
background object, then this system naturally falls into the prior
subclass.

As just mentioned, the next class of systems is the {\bf Minor
Mergers}. These are two (or more) galaxies which are physically close
to each other and show signs of interacting, and
in which one of the galaxies is much larger or much brighter than the
other(s). While we have by no means quantified the relative size or
luminosity ratio, we estimate that the companion is of order 1/4 or
less the size or luminosity of the primary. These systems are arranged in
order of decreasing separation between the component galaxies.
With the exception of M\,108, all the pairs of Minor Mergers
have known redshifts which imply a physical associaton.

Within the class of Minor Mergers, we include several subclasses. The
first is {\bf M\,51 Types}: large grand-design
spirals with small companions at the end of one of the spiral
arms. The next subclass is {\bf 3-body Encounters}, of which
M\,81 (Fig.~53) is the
prototype. These are different from the class of Triples later in the
Gallery, as one of the participants appears much larger or brighter
than the others. The Mrk\,348 system
(Fig.~52) is a member of both of these subclasses: it
has a small companion at the end of a spiral arm, and an \HI\
distribution very similar to M\,51 (although on a much larger physical
scale); but there is also a large neighbor
(NGC\,266) to the northwest, which may have
played some role in shaping the outer \HI\ morphology.

After the 3-body encounters comes NGC\,1097
(Fig.~57), which is placed here because its companion
(NGC\,1097A) appears about ready to merge with it. This is followed by
the {\bf Minor Merger Remnants}. These have a single identifiable
nucleus, but optical morphological peculiarities typically ascribed to
strong gravitational disturbances, such as shells, ripples, tails, and
plumes. They appear here, as opposed to the later class of {\it Merger 
Remnants}, because of the continued survival (or re-formation?) 
of a large disk. It is widely believed that major mergers destroy
disks. While we are not convinced of this fact, the literature on
these objects discusses them almost exclusively in terms of a minor
merger origin, so we adopt those results here.

The next class of galaxies is the dominant class represented in the
Gallery: {\bf Interacting Doubles or Major Mergers}. These are
galaxies of apparently similar mass which are physically
associated. This classification has taken precedence over other
classifications except {\it Triples --- Groups}. For example, if
the outer \HI\ morphology appears clearly tidal in origin, we
have placed the system in this class rather than under the
category of {\it \HI\ Extensions}.

We have subdivided this category into five separate subclasses. In
doing so we draw heavily on the lessons learned in the pioneering
study of interacting galaxies by Toomre \& Toomre (1972; see also
Barnes 1998). Specifically, from that work we learn the following:
(1)~tidal features first form shortly after first orbital periapses;
(2)~tails are formed at the rate of one per prograde disk;
(3)~high-inclination or retrograde encounters lead to large epicyclic
motions within the disk, but do not form well-defined tails; and
(4)~bridges form from a wide range of encounter geometries. Further, we
know that later-type spirals tend to be rich in \HI, whereas
lenticulars and ellipticals tend to be gas poor (Roberts \& Haynes
1994).

With these considerations in mind, we have morphologically defined
several subclasses which we think are dynamically related to the
Hubble types of the participants and the spin geometry of the
encounter. We emphasize that this classification is {\it purely
morphological}: we have made no attempt to check the proposed spin
geometries against the velocity fields of the galaxies.
The subclasses are defined as follows:

{\bf Two \HI\ Systems; Two Tails}: from the above dynamical
considerations, we suspect that these are interactions between two
prograde disk galaxies, which we indicate in Table~1 by the notation
($Sp^+ - Sp^+$), where $Sp$ denotes a spiral progenitor, and the $+$
superscript denotes the suggested prograde spin geometry. Objects
within this subclass are arranged by decreasing nuclear separation,
with well-separated objects at the beginning and single objects with
two tails at the end. II~Zw~40
(Fig.~6) may reasonably be placed midway through this
sequence, but we left it with the BCDs in the class of {\it Galaxies
with Extended \HI\ Envelopes}.

{\bf Two \HI\ Systems, One \HI\ Tail}: from the above dynamical
considerations, we suspect that these involve interactions between two
spiral disk galaxies, only one of which has a prograde geometry. The
second disk might have a highly inclined or a retrograde spin
geometry, indicated by a superscript $0$ in the ($Sp^+ - Sp^0$)
notation in Table~1. Objects within this subclass are again arranged
by decreasing nuclear separation, with well-separated objects at the
beginning and single objects with a single tail at the end.
Arp~295 (Fig.~159) belongs near the
beginning of this sequence, but has been placed with {\it Groups...}
due to its large number of \HI\ companions. The first member of this
subclass, the LMC/SMC/MS system (Fig.~73), may more
properly belong to the class of 3-body encounters under {\it Minor
Mergers}, but we decided to place it according to the morphology
exhibited in the figure.  Arp~215
(Fig.~60) might reasonably be placed at the end of
this sequence, rather than with the {\it Minor Merger Remnants}.

{\bf Two \HI\ Systems, Bridge, No Tails}: we suspect that these
involve interactions between two highly inclined or retrograde disk
galaxies, denoted by the ($Sp^0 - Sp^0$) notation in Table~1. The
first members of this subclass may well have different spin geometries
and be caught prior to first orbital periapse, but they fit the above
morphological definition so are placed here. The galaxies in this
subclass are arranged with nuclear separation first decreasing as
morphological distortion increases, then with nuclear separation
increasing as more bridge material appears between the two systems.
Notably, there are no Merger Remnants included at the end of this
sequence. Since the tidal signatures of such encounters are not as
well defined, it is much harder to make a confident classification of
such objects after the progenitors have merged. VV~114
(Fig.~107 under {\it Merger Remnants}) may be an
example of such an object.  The reader will notice that this sequence
ends with Ring Galaxies, which also starts the next subclass.

{\bf Two Systems, Only One with \HI}: These are presumably encounters
between one gas-rich and one gas-poor progenitor ($E - Sp$
notation). The first few systems in the subclass exhibit no tails,
while the remaining systems all show one tail. The first three
continue the theme of the last five systems of the previous subclass
by showing Ring Galaxies, but in this case the smaller penetrating
galaxy has no gas. It is possible that the gas was stripped as this
system passed through the gas-rich target system, or that the
progenitor was always gas-poor. In the Arp~104 system
(Fig.~101) it looks like the southern system has a
gas-rich tail, but the morphology of this feature resembles numerical
simulations in which bridge material passes through the companion and
emerges on the opposite side.

The final subclass of Interacting Doubles is that of {\bf Merger
Remnants of Indeterminate Origin}. These systems are clearly the
result of the coalescence of separate stellar systems, but it is
really not possible to say what has merged or how.

All of the previous classes have been dominated by late-type or spiral
galaxies. The next major class, {\bf Peculiar Early Types or Early
Types with Peculiar \HI}, is centered around early types and
ellipitcals.  Many of these may be Merger Remnants, others may be
Minor Merger Remnants, and yet others may owe their gas and/or optical
morphological peculiarities to their dense local environments. Since
these origins are very difficult to distinguish, we have made
subclasses based on both their optical and \HI\ morphology.

The first subclass is {\bf Peculiar Ellipticals with \HI\ Outside the
Optical Body}, and is ordered by the amount of \HI\ in the outer
regions (from lots of \HI\ to no \HI). This sequence may be considered
a possible extension of the Toomre Sequence of Major Mergers (Toomre
1977), demonstrating how gas-rich disk galaxies might fall together,
merge, and leave a gas-poor bulge-dominated galaxy in their place. The
amount of optical peculiarity decreases along this sequence (but not
uniformly), with more subtle optical peculiarities in the later than
in the earlier stages.  At the end of this sequence we have included a
montage of optically peculiar early types mapped in \HI, but in which
the \HI\ is not associated with the early type galaxy. These provide
an interesting counterpart to optically peculiar systems in which \HI\
has been detected. It is possible that more sensitive \HI\
observations might uncover some \HI\ in these systems, but clearly it
will be less than in those that have already been detected.  These
systems are ordered by decreasing optical peculiarity, as quantified
by the Fine Structure Index (FSI) of Schweizer \& Seitzer (1992). This
index quantifies the number of shells, jets, plumes, ripples, and
``X''-structures, as well as the boxiness of the galaxy. We note that
many of these galaxies exhibit quite striking peculiarities when
imaged with modern CCDs, and the reproductions here, taken from the
Digital Sky Survey, really do not do them full justice.

The next subclass is {\bf Peculiar Early Types with \HI\ Within the
Optical Body}. These are arranged by the degree of regularity of the
\HI. The early systems have a very irregular \HI\ distribution, and
the distribution becomes more symmetric and disk-like as the sequence
progresses. This sequence demonstrates the intriguing possibility
that in some cases enough cold gas is accreted onto a bulge-dominated
galaxy to form (or re-form) a disk.

This is followed by {\bf Normal Early Types with Peculiar
\HI}. These galaxies have no obvious optical peculiarities (certainly
at a much lower level than the preceding two classes), but have some
very interesting \HI\ distributions. This emphasizes the point that it
is very difficult, if not impossible, to guess the \HI\ morphology
based on a system's optical appearance.  The first system of this
subclass, the polar ring galaxy UGC\,7576
(Fig.~130), could equally well have been the last
system of the previous subclass. After this, the \HI\ distribution
becomes increasingly irregular along the sequence. Unlike the previous
two subclasses, this ordering is not meant to suggest an evolutionary
sequence. Rather, this ordering forms a natural transition from the
disk-like \HI\ distributions of the previous subclass, to the
irregular distribution of the next class of objects.

Given the normal optical appearance of the hosts, we are not sure what
to make of this subclass of objects. It is possible that future
optical observations will reveal as-yet-undiscovered optical
peculiarities in these systems, and that they may fit naturally into
one of the previous two categories. It is also feasible that the \HI\
has a tidal origin, and for some reason the encounter geometry left
the outer tidal gas in an irregular distribution long after the inner
regions have relaxed. In this case, these systems would be an
extension of the previous two categories. Yet another possibility is
suggested by the fact that most, if not all, of the members of this
subclass live in group environments; in this case the gas may have
been stripped from the outer regions of other members and accreted
onto the early type, which is usually one of the largest members of
the group. These objects may therefore belong to the class of {\it
Interacting Triples --- Groups --- Clusters}.  In the latter class
however the effects of interaction are manifested optically, so we
have elected to keep these classes separate.  The most interesting
possibility (in our opinion) is that the gas in these systems was
never in a galaxy, and represents accretion from a primordial
reservoir. At present it is not possible to discriminate between these
and other possible scenarios, but this is in any case one of the most
intriguing categories of objects in the Gallery.

The next class of objects is just as interesting, comprising galaxy
systems with {\bf Intergalactic Debris with No Optical Counterpart}. 
These three objects could easily have been categorized into other
classes (the Leo Ring with the previous class;
NGC\,5291 with peculiar early types with
\HI\ within the optical body; and NGC\,4532 with
galaxies with one-sided extensions, or with the minor mergers). We have
placed them into a separate category since in these cases the relation
between the \HI\ and the neighboring galaxies is less clear.

{\bf Interacting Triples --- Groups --- Clusters} form the next major
class.  As mentioned above, there are a lot of similarities between
the \HI\ distribution seen in triples and groups and those shown in
the previous three subclasses of objects, but here there are
clear optical distortions suggesting more directly an interaction
origin for the intergalactic gas.  This class is ordered (for the most
part) by the increasing number of members. The exception is at the end,
where we have put three early-type-dominated groups. In each of
these last three systems there is a significant extended X-ray
component, which must have some effect on the presence or absence of
cold gas.

Finally, there is the unavoidable {\bf Miscellaneous} class. These
objects do not obviously belong to any of the preceding classes, nor
are there enough similar characteristics to warrant the creation of
additional classes.  The first examples of this class (the
low-redshift QSOs and the E+A galaxy) have been called interacting
galaxies, but this conclusion was based on the \HI\ distribution. If
there is anything we have learned from the compilation of the Gallery,
it is that weird \HI\ distributions need not always arise due to 
interactions.

\section{Description of Gallery Figures}\label{sec:images}

We have attempted to present the data in as uniform a manner as
possible. To facilitate this, many users contributed the original data
in FITS format. The basic \HI\ data are the integrated \HI\ intensity
maps (zeroth moment), which give the integrated flux
($\int S \Delta v$; units of mJy\,beam$^{-1}$ km\,s$^{-1}$) at each
location. This is usually constructed with the windowing technique to
suppress the inclusion of noise (Bosma 1978).  The contour levels are given
in terms of the surface density, $N_{HI}$ (in units of H atoms per cm$^2$),
which is obtained from the integrated flux via the
equation (see Spitzer 1978, eqn.~3.38 ):

\begin{equation}
N_{HI} = {1.104\times 10^{21} {\rm cm}^{-2} \over \theta_x \times \theta_y}
{\int S \Delta v \over {\rm mJy\, beam^{-1}\, km\, s^{-1}}}
\end{equation}

\noindent where $\theta_x \times \theta_y$ is the full-width at
half-maximum (FWHM) size of the synthesized beam along the major
and minor axis, measured in arcseconds, and the gas is assumed to be
optically thin.

The $N_{HI}$ contour levels are given in each figure caption, and are
usually separated by factors of two.  For a small number of
systems, it was not possible to derive contour levels.

Spectral line observations also provide line-of-sight velocity
information. This information is very informative, but we have
decided against its inclusion here, mostly because we had our hands
full collecting the integrated intensity maps, but also because
such maps beg for a color reproduction. In the future we hope to make
the velocity and velocity dispersion maps available, but for the
present the readers should refer to the cited literature.

The \HI\ data is shown alongside and contoured upon optical images
of each object. When available, we use existing optical CCD data,
predominantly donated by the \HI\ observers. Where optical data were
not available, we use the Digitized Sky Survey (DSS) image obtained
from the Space Telescope Science Institute. We use the second
generation survey products when available, prefering the blue plates
to the red.

There is a wealth of information available in the data, and a single
format does not do it justice. We have therefore used a combination of
layouts. In order to reproduce the figures at as large a scale as
possible, the figure captions do not describe the figure layout, which
we believe to be reasonably self-evident. The various layouts are
described here.  The simplest involves a greyscale representation of
the optical image on the left, and the optical image with \HI\ column
density contours on the right. When the faint optical structure
warrants it, we present two representations of the optical data with
different transfer functions, in addition to the optical image with
\HI\ contours superimposed. Finally, when the complexity of the \HI\
structure warrants it, we also present a greyscale representation of
the integrated \HI\ emission. These images show clearly \HI\ minima
and maxima, which may be ambiguous in the contour maps. When we do
show a greyscale map of the \HI, we sometimes include contours. These
are most often single contours from the optical data. When there are
multiple contours, these are spaced by factors of ten apart. On a few
occasions we show \HI\ contours upon the \HI\ greyscales. Since the
\HI\ and optical morphologies are so different, it should be obvious
what the contours represent.

Positive \HI\ contours are drawn with dark solid lines, and negative
contours (e.g., \HI\ absorption against radio continuum sources) are drawn
with dashed and/or light lines. Occasionally a larger or smaller field of
view (FOV) is also shown, to illustrate either the large-scale \HI\
distribution, or more distant \HI-detected companions, or to show 
details of the inner regions. The smaller FOV is usually indicated by 
dotted boxes in the larger FOV image.

As mentioned above, the information in the figure captions is kept to
a minimum. For the \HI\ data we report the telescope (and array
configuration for VLA data), synthesized beam resolution (FWHM), and
contour level. For the optical data we report the telescope and filter
combination, or simply ``DSS'' when data from the Digital Sky Survey
are used, and we give for each image the approximate FOV in arcminutes.
There is a section for minimal notes, and finally the
reference. Unpublished data is listed as ``[authors], these
proceedings'' with a page number. These refer either to contributions
presented at the conference and collected in the first part of these
proceedings, or to abstracts collected at the end of the Gallery.

In the first panel of each figure we label each galaxy with a catalog
designation and the Hubble Type reported in the NASA Extragalactic
Database (NED). We also indicate whether NED lists the galaxy with a
Seyfert classification. These classifications have not been verified,
and should only be used as rough guides. They are included because we find
it interesting when galaxies classified as early types are found to
have lots of \HI, and when galaxies classified as spirals are found to
have none or very little. Using NED we have attempted to identify all
cataloged galaxies with \HI\ detections, as well as galaxies with
known redshifts which place them within the range of the \HI\
observations. Galaxies with their names labeled in parenthesis have
redshifts which put them in the foreground or background.

\section{Tables}\label{sec:tables}

At the beginning of the Gallery we present three tables.

Table~1 outlines the arrangement of the Gallery
by morphological class and subclass as described in \S3. It gives
one line descriptions of the classes and the abbreviations
used to indicate the class in Table 3.

Table~2 is the Table of Contents for the Rogues Gallery. It lists the
galaxies in the order in which they appear in the Atlas, grouped by
class, followed by the Figure and page number.

Table~3 presents basic information for all 432 of the cataloged
galaxies labeled in the 181 images of the Rogues Gallery. The entries
are listed in order of increasing Right Ascension (J2000) of the main
target galaxy in the image. The galaxies are grouped by image and
groups are separated by thick horizontal lines. Within a group
individual galaxies are separated by thin horizontal lines. In the
first two columns we list the most popular catalog names. In column
three we list the J2000 coordinates and the heliocentric radial
velocity (km sec$^{-1}$).  All information in these first three
columns is taken directly from a batch query to the NASA Extragalactic
Database (NED) in July of 2001. Column 4 lists the morphological type
as found in NED, the RSA (Sandage \& Tamman 1987) and the RC3 catalogs
(de Vaucouleurs et al.  1991). Column 5 gives the class as defined in
\S3 and Table 1, and the Figure and page number.

In addition to the three table mentioned above, there is a
comprehensive object index at the end of this volume, which includes
all objects in the Rogues Gallery.

\section{Acknowledgments}
\label{sec:acknowledgements}

We could not have compiled this collection without the generosity of
dozens of \HI\ observers within the community. Many of these
contributions came as part of the general conference, although others
who were unable to attend also contributed. We would like to offer our
sincere thanks to these people for making this gallery possible, and
implore the reader to properly reference the original
work. Specifically, this atlas has greatly benefited from
the specific contributions by
Tyler Nordgren, Marc Verheijen, Jim Higdon, Rob Swaters, Caroline
Simpson and collaborators, Jeremy Lim, D.J.\ Pisano, Marcel Clemens,
Carole Mundell, Salman Hameed, Eric Wilcots, Liese van Zee,
Pierre-Alain Duc, Dave Hogg, Judith Irwin, Michele Kaufman, Bev Smith,
Veera Boonyasait, Elias Brinks, Vassilis Charmandaris, Jayanne
English, Deidre Hunter, Linda Sparke, Andrea Cox, Lourdes
Verdes-Montenegro, Min Yun, Phil Appleton, Jim Condon, Michiel Kregel,
Glen Langston, Tom Oosterloo, Oak-Kyoung Park, Mary Putman, Rich Rand,
Arnold Rots, Sue Simkin, Lister Staveley-Smith, Athanasios
Taramopoulos, Helen Thomas, Wei-Hao Wang, and Barbara Williams. We
also thank Garrett Bauer and Karen Yeh for adding the FOVs to the
image captions.

The VLA is a facility of the National Radio Astronomy Observatory,
which is operated by Associated Universities Inc. under cooperative
agreement with the National Science Foundation.

The Digitized Sky Surveys were produced at the Space Telescope Science
Institute under U.S. Government grant NAG W-2166. The images of these
surveys are based on photographic data obtained using the Oschin
Schmidt Telescope on Palomar Mountain and the UK Schmidt
Telescope. The Oschin Schmidt Telescope is operated by the California
Institute of Technology and Palomar Observatory.  The UK Schmidt
Telescope was operated by the Royal Observatory Edinburgh, with
funding from the UK Science and Engineering Research Council (later
the UK Particle Physics and Astronomy Research Council), until 1988
June, and thereafter by the Anglo-Australian Observatory.  The plates
were processed into the present compressed digital form with the
permission of these institutions.

The National Geographic Society - Palomar Observatory Sky Atlas was
made by the California Institute of Technology with grants from the
National Geographic Society.  The Second Palomar Observatory Sky
Survey was made by the California Institute of Technology with funds
from the National Science Foundation, the National Geographic Society,
the Sloan Foundation, the Samuel Oschin Foundation, and the Eastman
Kodak Corporation.

This research has made extensive use of the NASA/IPAC Extragalactic
Database (NED) which is operated by the Jet Propulsion Laboratory,
California Institute of Technology, under contract with the National
Aeronautics and Space Administration.

\vfill\newpage

\end{document}